\documentclass{iopart}
\eqnobysec

\begin{document}
\title{An integrable discretization of KdV at large times}

\author{M Boiti\dag, F Pempinelli\dag, B Prinari\dag and A. Spire\ddag}
\address{\dag\ Dipartimento di Fisica dell'Universit\`{a} and Sezione INFN,
73100 Lecce, Italy}
\address{\ddag\ Physique Math\'{e}matique et Th\'{e}orique, CNRS-UMR5825,
Universit\'{e} Montpellier 2, 34095 Montpellier, France}

\begin{abstract}
An ``exact discretization'' of the Schr\"{o}dinger operator is considered
and its direct and inverse scattering problems are solved. It is shown that
a differential-difference nonlinear evolution equation depending on two
arbitrary constants can be solved by using this spectral transform and that
for a special choice of the constants it can be considered an integrable
discretization of the KdV equation at large times. An integrable difference-difference
equation is also obtained.
\end{abstract}

\maketitle

\section{Introduction}

We are interested in discrete dynamical systems that in the continuous limit
give the KdV equation and that are integrable by means of the inverse
scattering technique.

It is long time that the inverse scattering method was extended to solve
differential-difference and difference-difference nonlinear equations. See
in particular the pioneer works of Flaschka \cite{Flaschka}, Manakov \cite
{ManakovDiscrete}, Ablowitz and Ladik \cite{Ablowitz-Ladik1,Ablowitz-Ladik2},
Kac and van Moerbeke \cite{KacMoerbeke}. In these papers and in following
papers (among other discrete equations) different versions of the discrete
KdV equation were introduced and studied. See \cite{Hirota}-\cite
{Bogoyavlensky}, the review paper \cite{NijhoffReview} and references quoted
therein.

The various already considered discrete KdV equations were related to a
reduced form of the Ablowitz-Ladik spectral problem or to different
discretized versions of the Schr\"{o}dinger spectral problem. See, in
addition, for the spectral theory of these discrete Schr\"{o}dinger
operators \cite{Case}--\cite{Guseinov2} and \cite
{Bruschiautomata,Pogrebautomata} for the case of cellular automata.

Here we use the ``exact discretization'' of the Schr\"{o}dinger equation
considered recently by Shabat
\begin{equation}
\psi _{n+2}=g_{n}\psi _{n+1}+\lambda \psi _{n}  \label{Shabat}
\end{equation}
and obtained iterating the Darboux transformations \cite{Shabat}.

We completely solve the direct and inverse scattering problems in the case
of a potential $g_{n}$ satisfying
\begin{equation}
\sum_{n=-\infty }^{+\infty }(1+|n|)|g_{n-1}-2|<\infty .  \label{gcondition}
\end{equation}
Then we associate to the discrete spectral problem (\ref{Shabat}) the
differential-difference equation ($p$ and $q$ are arbitrary constant)
\begin{equation}
\left( g_{n}g_{n-1}\right) _{t}=p\left( \frac{g_{n-1}}{g_{n+1}}-\frac{g_{n}}{g_{n-2}}
\right) +q\left( g_{n}-g_{n-1}\right)g_{n}g_{n-1}  \label{equationg}
\end{equation}
and we solve its initial value Cauchy problem in the standard way by using
the inverse scattering method.

Different possible continuous limits remain to be explored. In the special
case
\begin{equation}
p=2q
\end{equation}
if we let
\begin{eqnarray}
  g_{n}=2+h^{2}u(nh) \\
  x=nh,\quad T=h^{3}\frac{p}{8}t
\end{eqnarray}
in the limit $h\rightarrow 0$ we get the KdV equation
\begin{equation}
u_{T}=-u_{xxx}+6u_{x}u.
\end{equation}
Therefore the equation (\ref{equationg}), in the case $p=2q$, can be
considered an integrable discretization at large times of the KdV equation.

Finally, we consider also a double discrete equation, i.e. discrete both in
space and in time, associated to the double discrete Schr\"{o}dinger
equation
\begin{equation}
\psi _{n+2,m}=g_{n,m}\psi _{n+1,m}+\lambda \psi _{n,m},
\end{equation}
that in the limit of continuous time reduces to (\ref{equationg}).

\section{Differential-difference equation}

In the following it is sometimes convenient to use, for any discrete function $f(n)$,
the notation
\begin{equation}
f_{i}=f\left( n+i\right) .
\end{equation}
when $n$ is considered generic but fixed.

Then the discrete Schr\"{o}dinger equation (\ref{Shabat}) reads
\begin{equation}
\psi _{2}=g_{0}\psi _{1}+\lambda \psi _{0}.  \label{Shabatshort}
\end{equation}

We consider the auxiliary spectral problem
\begin{equation}
\psi _{1t}=\left( A_{0}+\lambda B_{0}\right) \psi _{0}+(C_{1}+\lambda
D_{1})\psi _{1},  \label{6}
\end{equation}
and we impose compatibility in order to determine the coefficients in the
r.h.s. and the nonlinear differential-difference evolution equation which
can be solved by using the spectral transform related to (\ref{Shabatshort}).

Precisely, let us denote by $E$ the shift operator, $Ev_{n}=v_{n+1}$. Then,
the compatibility condition between (\ref{Shabatshort}) and (\ref{6}), that
is $E\psi _{1t}=\partial _{t}\psi _{2}$, gives
\begin{eqnarray}
\fl \left( A_{1}+\lambda B_{1}\right) \psi _{1}+\lambda \left( C_{2}+\lambda
D_{2}\right) \psi _{0}+\left( C_{2}+\lambda D_{2}\right) g_{0}\psi _{1} \nonumber\\
{ }=\left( A_{-1}+\lambda B_{-1}\right) \psi _{1}-(A_{-1}+\lambda
B_{-1})g_{-1}\psi _{0}+\lambda (C_{0}+\lambda D_{0})\psi _{0} \nonumber\\
\qquad\qquad { }+g_{0t}\psi _{1}+(A_{0}+\lambda B_{0})g_{0}\psi _{0}+(C_{1}+\lambda
D_{1})g_{0}\psi _{1}. \nonumber
\end{eqnarray}
Assuming $\psi _{1}$ and $\psi _{0}$ to be linearly independent and equating
the coefficients of the different powers in $\lambda $ we get the following
relations
\begin{eqnarray}
  D_{2}=D_{0}  \label{11} \\
  C_{2}-C_{0}=g_{0}B_{0}-g_{-1}B_{-1}  \label{12} \\
  A_{0}g_{0}=A_{-1}g_{-1}  \label{13} \\
  B_{1}-B_{-1}=g_{0}(D_{1}-D_{2})  \label{14} \\
  g_{0t}=A_{1}-A_{-1}+g_{0}\left( C_{2}-C_{1}\right) .  \label{15}
\end{eqnarray}
From (\ref{11}), \ref{13}) and (\ref{14}) we immediately have
\begin{eqnarray}
  D(n)=d  \label{D} \\
  A(n)g(n)=p  \label{16} \\
  B(n)=q
\end{eqnarray}
where $p$, $q$ and $d$ are constants, and then we are left with
\begin{equation}
C_{2}-C_{0}=q\left( g_{0}-g_{-1}\right)  \label{17}
\end{equation}
and
\begin{equation}
g_{0t}=p\left( \frac{1}{g_{1}}-\frac{1}{g_{-1}}\right) +g_{0}(C_{2}-C_{1}).
\label{18}
\end{equation}
We write (\ref{18}) for the shifted indices
\begin{equation}
g_{-1t}=p\left( \frac{1}{g_{0}}-\frac{1}{g_{-2}}\right) +g_{-1}(C_{1}-C_{0})
\label{18'}
\end{equation}
and sum it to (\ref{18}). Using (\ref{17}) we finally get the searched
differential-difference equation
\begin{equation}
\left( g_{0}g_{-1}\right) _{t}=p\left( \frac{g_{-1}}{g_{1}}-
\frac{g_{0}}{g_{-2}}\right) +q\left( g_{0}-g_{-1}\right)g_{0}g_{-1} .  \label{19}
\end{equation}
The coefficients of the auxiliary spectral problem are given by
\begin{eqnarray}
A(n)  = \frac{p}{g(n)}  \label{A0} \\
B(n)  = q  \label{B0} \\
C(n)  = q\sum_{j=1}^{\infty }\left( g(n-2j)-g(n-2j-1)\right)  \label{C1} \\
D(n)  = 0,  \label{D1}
\end{eqnarray}
where for convenience we chose equal to zero the integration constant of $C$
and $D$. In the following we need the limit of these coefficients as
$n\rightarrow \pm \infty $. For $n\rightarrow -\infty $ we have $C(n)\rightarrow
0$, but for $n\rightarrow +\infty $ it is necessary to
consider separately the cases odd $n$ and even $n$. We have
\begin{eqnarray}
C(2m)  = q\sum_{r=-\infty }^{m-1}\left( g(2r)-g(2r-1)\right) \\
C(2m+1)  = q\sum_{r=-\infty }^{m-1}\left( g(2r+1)-g(2r)\right) .
\end{eqnarray}
The limits
\begin{eqnarray}
c_{e}  = \lim_{m\rightarrow +\infty }C(2m) \\
c_{d}  = \lim_{m\rightarrow +\infty }C(2m+1),
\end{eqnarray}
thanks to (\ref{gcondition}) exist and satisfy
\begin{equation}
c_{e}+c_{d}=0.  \label{cecd0}
\end{equation}

\section{Spectral transform}

\subsection{Direct problem}

The spectral problem (\ref{Shabat}) by using notations
\begin{eqnarray}
  \lambda =-1-k^{2}  \label{lambdak} \\
  g(n)=2+u(n) \\
  \psi (n;k)=(1+ik)^{n}\chi (n;k)  \label{psichi}
\end{eqnarray}
can be more conveniently rewritten as
\begin{equation}
\fl (1+ik)\chi (n+2;k)-2\chi (n+1;k)+(1-ik)\chi (n;k)=u(n)\chi (n+1;k).
\label{mod-eq}
\end{equation}

The Jost solutions can be defined via the following discrete integral
equations
\begin{eqnarray}
\mu ^{+}(n;k)  = 1-\frac{1}{2ik}\sum_{j=n+1}^{+\infty }\left[ 1-\left(
\frac{1+ik}{1-ik}\right) ^{j-n}\right] u(j-1)\mu ^{+}(j;k)  \label{int-mu} \\
\mu ^{-}(n;k)  = 1+\frac{1}{2ik}\sum_{j=-\infty }^{n}\left[ 1-\left(
\frac{1+ik}{1-ik}\right) ^{j-n}\right] u(j-1)\mu ^{-}(j;k).  \label{int-nu}
\end{eqnarray}
We show in the next section that for a potential $u(n)$ decaying
sufficiently rapidly to zero at large $n$ the Jost solutions $\mu ^{+}$ and
$\mu ^{-}$ are, respectively, analytic functions of $k$ in the upper and in
the lower half-plane and have a continuous limit on the real $k$ axis. Then,
it is natural to introduce the following spectral data (defined on the real
axis $k_{\Im }=0$)
\begin{eqnarray}
a^{\pm }(k)  = 1\mp \frac{1}{2ik}\sum_{j=-\infty }^{+\infty }u(j-1)\mu ^{\pm
}(j;k)  \label{a} \\
b^{\pm }(k)  = \pm \frac{1}{2ik}\sum_{j=-\infty }^{+\infty }\left( \frac{1+ik
}{1-ik}\right) ^{j}u(j-1)\mu ^{\pm }(j;k)  \label{b} \\
\rho ^{\pm }(k)  = \frac{b^{\pm }(k)}{a^{\pm }(k)}.  \label{rho-rhobar}
\end{eqnarray}
Due to the above mentioned properties of analyticity of $\mu ^{\pm }$ , it
is clear that $a^{\pm }$ can be analytically extended, respectively, to the
upper half-plane and to the lower half-plane of the complex spectral
parameter. The function $1/a$ plays the role of transmission coefficient
and $\rho $ is the reflection coefficient.

By using the definitions of $\mu ^{\pm }$ in (\ref{int-mu}) and (\ref{int-nu})
one can check directly that on the real $k$-axis, where both Jost
solutions are defined,
\begin{eqnarray}
\fl \left[ \frac{\mu ^{+}(n;k)}{a^{+}(k)}-\mu ^{-}(n;k)\right]
\left( \frac{1+ik}{1-ik}\right) ^{n}=\rho ^{+}(k) \nonumber \\
\fl \qquad\quad{ }-\frac{1}{2ik}\sum_{j=-\infty }^{n}\left[ 1-\left( \frac{1+ik}{1-ik}\right)
^{n-j}\right] u(j-1)\left( \frac{1+ik}{1-ik}\right) ^{j}\left[ \frac{\mu
^{+}(j;k)}{a^{+}(k)}-\mu ^{-}(j;k)\right] \nonumber
\end{eqnarray}
so that, assuming that the ``integral equation'' (\ref{int-nu}) is uniquely
solvable, we get
\begin{equation}
\frac{\mu ^{+}(n;k)}{a^{+}(k)}=\mu ^{-}(n;k)+\left( \frac{1-ik}{1+ik}\right)
^{n}\rho ^{+}(k)\mu ^{-}(n;-k),\quad k=k_{\Re }  \label{spectr1}
\end{equation}
Analogously we obtain
\begin{equation}
\frac{\mu ^{-}(n;k)}{a^{-}(k)}=\mu ^{+}(n;k)+\left( \frac{1-ik}{1+ik}\right)
^{n}\rho ^{-}(k)\mu ^{+}(n;-k),\quad k=k_{\Re }.  \label{spectr2}
\end{equation}
Taking into account the integral equations for $\mu ^{\pm }$, we see that
\begin{eqnarray}
\lim_{n\rightarrow +\infty }\mu ^{+}(n;k)  = 1,\qquad k_{\Im }\geq 0,  \label{nu-} \\
\lim_{n\rightarrow -\infty }\mu ^{-}(n;k)  = 1,\qquad k_{\Im
}\leq 0  \label{nu-2}
\end{eqnarray}
while for $k_{\Im }=0$ we have
\begin{equation}
\mu ^{\pm }(n;k)\simeq a^{\pm }(k)+\left( \frac{1-ik}{1+ik}\right)
^{n}b^{\pm }(k),\quad \quad {\rm for }\;\; n\rightarrow \mp \infty .  \label{nu+}
\end{equation}
Note that $\left( \frac{1-ik}{1+ik}\right) ^{n}$ gives rise to an
oscillating term, since $\left| \frac{1-ik}{1+ik}\right| =1$ for $k=k_{\Re }$.
Then, from the asymptotics at large $n$ of (\ref{spectr1}) and (\ref
{spectr2}) we get the unitarity relations ($k=k_{\Re }$)
\begin{eqnarray}
  a^{+}(k)a^{-}(k)+b^{+}(k)b^{-}(-k)=1 \\
  b^{+}(k)b^{-}(-k)=b^{+}(-k)b^{-}(k) \\
  a^{+}(k)b^{-}(k)+b^{+}(k)a^{-}(-k)=0.
\end{eqnarray}
It is convenient to introduce also
\begin{eqnarray}
\nu ^{+}(n;k)  = \left( \frac{1-ik}{1+ik}\right) ^{n}\mu ^{-}(n;-k) \\
\nu ^{-}(n;k)  = \left( \frac{1-ik}{1+ik}\right) ^{n}\mu ^{+}(n;-k)
\end{eqnarray}
which are solutions of the spectral problem (\ref{mod-eq}) and, up to the
kinematical factor $\left( \frac{1-ik}{1+ik}\right) ^{n}$, are,
respectively, analytic in the upper and lower half-plane of the spectral
parameter.

If one defines the discrete Wronskian as
\begin{equation}
\fl W(\chi ,\varphi )(n)=\left( \frac{1+ik}{1-ik}\right) ^{n}\left[ \chi
(n+1;k)\varphi (n;k)-\chi (n;k)\varphi (n+1;k)\right]
\end{equation}
it is easy to verify that for two solutions $\chi $ and $\varphi $ of (\ref
{mod-eq}) the Wronskian is independent of $n$. Then, we can evaluate $W(\mu
^{+},\nu ^{+})$ using the asymptotic limits of $\mu ^{+}(n;k)$ and $\mu
^{-}(n;-k)$ for $n\rightarrow -\infty $ in (\ref{nu-}), (\ref{nu-2}) and
(\ref{nu+}) getting
\begin{equation}
W(\mu ^{+},\nu ^{+})(k)=\frac{2ik}{1+ik}a^{+}(k).  \label{Wa+}
\end{equation}
Of special interest are the zeros of $a^{+}(k)$, say at $k=k_{m}^{+}$ ($
m=1,2,\dots ,N^{+}$), which are located in the upper half-plane and that we
expect to be related to the soliton solutions of the evolution equations
related to the spectral problem (\ref{mod-eq}). From (\ref{Wa+}) we deduce
that $\mu ^{+}(n;k_{m}^{+})$ and $\nu ^{+}(n;k_{m}^{+})$ are proportional,
i.e.
\begin{equation}
\mu ^{+}(n;k_{m}^{+})=D_{m}^{+}\nu ^{+}(n;k_{m}^{+}).  \label{muDnu}
\end{equation}
The constant $D_{m}^{+}$ can be deduced by rewriting the integral equation
(\ref{int-mu}) for $\mu ^{+}$ at $k=k_{m}^{+}$ as follows
\begin{eqnarray}
\fl \left( \frac{1+ik_{m}^{+}}{1-ik_{m}^{+}}\right) ^{n}\mu ^{+}(n;k_{m}^{+})=
\frac{1}{2ik_{m}^{+}}\sum_{j=-\infty }^{+\infty }\left( \frac{1+ik_{m}^{+}}{
1-ik_{m}^{+}}\right) ^{j}u(j-1)\mu ^{+}(j;k_{m}^{+}) \nonumber \\
\fl \qquad \qquad { }-\frac{1}{2ik_{m}^{+}}\sum_{j=-\infty }^{n}\left[ 1-\left(
\frac{1-ik_{m}^{+}}{1+ik_{m}^{+}}\right)^{j-n}\right] u(j-1)
\left( \frac{1+ik_{m}^{+}}{
1-ik_{m}^{+}}\right) ^{j}\mu ^{+}(j;k_{m}^{+}).\nonumber
\end{eqnarray}
Comparing with the integral equation (\ref{int-nu}) defining $\mu ^{-}$ we
get
\begin{equation}
D_{m}^{+}=\frac{1}{2ik_{m}^{+}}\sum_{j=-\infty }^{+\infty }\left(
\frac{1+ik_{m}^{+}}{1-ik_{m}^{+}}\right) ^{j}u(j-1)\mu ^{+}(j;k_{m}^{+}).
\label{Dm}
\end{equation}
Notice that the convergence of the series is guaranteed by the
proportionality between $\mu ^{+}(n;k_{m}^{+})$ and $\nu ^{+}(n;k_{m}^{+})$
that have, respectively, good behaviour for $n\rightarrow +\infty $ and $
n\rightarrow -\infty $.

\subsection{Neumann series for $\protect\mu ^{\pm }$}

We consider the series
\begin{equation}
\mu ^{+}(n;k)=\sum_{l=0}^{+\infty }\mu ^{(l)}(n;k)  \label{mu+series}
\end{equation}
whose coefficients are defined through the iterative algorithm
\begin{equation}
\mu ^{(l+1)}(n;k)=\sum_{j=n+1}^{+\infty }M(n,j;k)u(j-1)\mu ^{(l)}(j;k)
\label{Neumann-mu}
\end{equation}
starting at $\mu ^{(0)}(n;k)=1$, where
\begin{equation}
M(n,j;k)=-\frac{1}{2ik}\left[ 1-\left( \frac{1+ik}{1-ik}\right) ^{j-n}\right]
.
\end{equation}
If the series is uniformly convergent it solves the discrete integral
equation (\ref{int-mu}) defining the Jost solution $\mu ^{+}$. This study is
most conveniently done following the procedure outlined in the continuous
case in \cite{Sabatier}.

We have for $j\geq n+1$ and $k_{\Im }\geq 0$, at the same time,
\begin{eqnarray}
\left| M(n,j;k)\right|  \leq  \frac{1}{2\left| k\right| }\left[ 1+\left|
\frac{1+ik}{1-ik}\right| ^{j-n}\right] \leq \frac{1}{\left| k\right| } \nonumber \\
\left| M(n,j;k)\right|  \leq  \left| \frac{1}{1-ik}\right|
\sum_{r=0}^{j-n-1}\left| \frac{1+ik}{1-ik}\right| ^{r}\leq (j-n).\nonumber
\end{eqnarray}
Now setting
\begin{equation}
I(x)=1+|x|\theta (-x)
\end{equation}
and noticing that for $(j-n)\geq 0$
\begin{equation}
(j-n)\leq I(n)I(-j)
\end{equation}
we can write
\begin{equation}
\left| M(n,j;k)u(j-1)\right| \leq \alpha (n)\beta (j)
\end{equation}
where
\begin{eqnarray}
\alpha (n)  = \left\{
\begin{array}{c}
|k|^{-1} \\
I(n)
\end{array}
\right. \\
\beta (j)  = |u(j-1)|\left\{
\begin{array}{c}
1 \\
I(-j)
\end{array}
\right. .
\end{eqnarray}
Notice that
\begin{equation}
\alpha (j)\beta (j)=|u(j-1)|\left\{
\begin{array}{c}
|k|^{-1} \\
(1+|j|)
\end{array}
\right. .
\end{equation}
Using these bounds in (\ref{Neumann-mu}) we obtain as majorant $\widehat{\mu
}^{(l)}$ of $|\mu ^{(l)}|$ the corresponding term of a series expansion of
the solution of
\begin{equation}
\widehat{\mu }(n;k)=1+\sum_{j=n+1}^{+\infty }\alpha (n)\beta (j)\widehat{\mu
}(j;k).
\end{equation}
One can prove by induction that for $l\geq 0$
\begin{equation}
\widehat{\mu }^{(l+1)}(n)\leq \sum_{j=n+l+1}^{+\infty }\alpha (n)\beta (j)
\frac{Q^{l}(n+1,j-1)}{l!}  \label{bound-mu}
\end{equation}
where
\begin{equation}
Q(n,j)=\sum_{s=n}^{j}\alpha (s)\beta (s).
\end{equation}
In fact, if (\ref{bound-mu}) is true for $l$, we have
\begin{equation*}
\widehat{\mu }^{(l+1)}(n)\leq \sum_{j=n+1}^{+\infty }\alpha (n)\beta
(j)\sum_{r=j+l}^{+\infty }\alpha (j)\beta (r)\frac{Q^{l-1}(j+1,r-1)}{(l-1)!}.
\end{equation*}
Changing the order of the sums we obtain
\begin{equation*}
\widehat{\mu }^{(l+1)}(n)\leq \sum_{r=n+l+1}^{+\infty }\alpha (n)\beta
(r)\sum_{j=n+1}^{r-l}\alpha (j)\beta (j)\frac{Q^{l-1}(j+1,r-1)}{(l-1)!}
\end{equation*}
and then, remarking that, since $Q(j,r-1)\geq Q(j+1,r-1)$,
\begin{eqnarray}
\fl Q^{l}(j,r-1)\geq Q^{l}(j,r-1)-Q^{l}(j+1,r-1) \nonumber \\
\fl \qquad{ }=(Q(j,r-1)-Q(j+1,r-1))\sum_{s=0}^{l-1}Q^{l-1-s}(j,r-1)Q^{s}(j+1,r-1)\geq \nonumber \\
\fl \qquad\qquad{ }\geq l\alpha (j)\beta (j)Q^{l-1}(j+1,r-1) \nonumber
\end{eqnarray}
we derive (\ref{bound-mu}).

Then, we change the summation limit in (\ref{bound-mu}) as follows
\begin{equation}
\widehat{\mu }^{(l+1)}(n)\leq \sum_{j=n}^{+\infty }\alpha (n)\beta (j)\frac{
Q^{l}(n+1,j-1)}{l!}
\end{equation}
and sum up over $l$ getting
\begin{equation}
\widehat{\mu }(n)\leq 1+\sum_{j=n}^{+\infty }\alpha (n)\beta (j)\exp
\left[Q(n+1,j-1)\right]
\end{equation}
and finally
\begin{equation}
\widehat{\mu }(n)\leq 1+C\sum_{j=n}^{+\infty }|u(j-1)|\left\{
\begin{array}{c}
|k|^{-1} \\
I(-j)I(n)
\end{array}
\right. ,
\end{equation}
where
\begin{equation}
C=\exp \left( \sum_{j=-\infty }^{+\infty }(1+|j|)|u(j-1)|\right) .
\end{equation}

Thus for $k_{\Im }\geq 0$ and for a potential $u$ satisfying
\begin{equation}
\sum_{j=-\infty }^{+\infty }(1+|j|)|u(j-1)|<\infty
\end{equation}
the series in the r.h.s. of (\ref{mu+series}) is uniformly bounded and each
iterated term $\mu ^{(l)}$ is absolutely bounded by the $l$-th term of a
convergent series. Since the $\mu ^{(l)}(n;k)$'s are analytic functions of
$k $ for $k_{\Im }>0$, we conclude that $\mu ^{+}(n;k)$ is analytic for $
k_{\Im }>0$ and has a continuous limit as $k_{\Im }\rightarrow 0$. A quite
similar analysis shows that $\mu ^{-}(n;k)$ is analytic for $k_{\Im }<0$ and
has a continuous limit as $k_{\Im }\rightarrow 0$.

\subsection{Inverse Problem}

Let us suppose that $a^{+}$ has simple zeros at $k_{j}^{+}$ and let us
denote by $c_{j}^{+}$ the residuum of $1/a^{+}$, that is
\begin{equation}
c_{j}^{+}=\lim_{k\rightarrow k_{j}^{+}}\frac{(k-k_{j}^{+})}{a^{+}(k)}\qquad
j=1,\dots ,N^{+}.  \label{residuals}
\end{equation}
Then we can write (\ref{spectr1}) in the form ($k=k_{\Re }$)
\begin{eqnarray}
\fl \left[ \frac{\mu ^{+}(n;k)}{a^{+}(k)}-\sum_{j=1}^{N^{+}}c_{j}^{+}\frac{\mu
^{+}(n;k_{j}^{+})}{k-k_{j}^{+}}\right] -\left[ \mu
^{-}(n;k)-\sum_{j=1}^{N^{+}}c_{j}^{+}\frac{\mu ^{+}(n;k_{j}^{+})}{k-k_{j}^{+}}\right]
\nonumber \\
{ }=\mu ^{-}(n;-k)\rho ^{+}(k)\left( \frac{1-ik}{1+ik}\right) ^{n}.\nonumber
\end{eqnarray}
The terms in the left-hand side are the limiting values at the real axis of
a sectionally holomorphic function; then, taking into account the asymptotic
behavior of $\mu^{\pm}$ at large $k$ and recalling (\ref{muDnu}), we can use the
Cauchy-Green formula and get for $k_{\Im }\geq 0$
\begin{eqnarray}
\fl \frac{\mu ^{+}(n;k)}{a^{+}(k)}=1+\sum_{j=1}^{N^{+}}C_{j}^{+}\frac{\mu
^{-}(n;-k_{j}^{+})}{k-k_{j}^{+}}\left( \frac{1-ik_{j}^{+}}{1+ik_{j}^{+}}
\right) ^{n}\nonumber \\
{ }+\frac{1}{2i\pi }\int_{-\infty }^{+\infty }\frac{\mu ^{-}(n;-s)\rho ^{+}(s)}{
s-k-i0}\left( \frac{1-is}{1+is}\right) ^{n}ds  \label{Cauchy-mu}
\end{eqnarray}
and for $k_{\Im }\leq 0$
\begin{eqnarray}
\fl \mu ^{-}(n;k)=1+\sum_{j=1}^{N^{+}}C_{j}^{+}\frac{\mu ^{-}(n;-k_{j}^{+})}{
k-k_{j}^{+}}\left( \frac{1-ik_{j}^{+}}{1+ik_{j}^{+}}\right) ^{n} \nonumber \\
{ }+\frac{1}{2i\pi }\int_{-\infty }^{+\infty }\frac{\mu ^{-}(n;-s)\rho ^{+}(s)}{
s-k+i0}\left( \frac{1-is}{1+is}\right) ^{n}ds  \label{Cauchymu-}
\end{eqnarray}
where
\begin{equation}
C_{j}^{+}=c_{j}^{+}D_{j}^{+}.
\end{equation}

Equations (\ref{Cauchy-mu}) and (\ref{Cauchymu-}) allow to reconstruct the
Jost solutions, given the spectral data. In order to complete the
formulation of the inverse problem, we have to reconstruct the potential $u$.
By inserting the asymptotic behaviour of $\mu ^{-}$ at large $k$
\begin{equation}
\mu ^{-}(n;k)=1+\frac{\mu _{(-1)}^{-}(n)}{k}+o(k^{-1})  \label{exp-mu}
\end{equation}
into (\ref{mod-eq}) we get
\begin{equation}
u(n)=i\left[ \mu _{(-1)}^{-}(n+2)-\mu _{(-1)}^{-}(n)\right] .  \label{rec-g}
\end{equation}
Therefore from (\ref{Cauchymu-}) we have
\begin{eqnarray}
\fl u(n)=i\sum_{j=1}^{N^{+}}\left[ \mu ^{-}(n+2;-k_{j}^{+})\left( \frac{1-ik_{j}^{+}
}{1+ik_{j}^{+}}\right) ^{2}-\mu ^{-}(n;-k_{j}^{+})\right] \left( \frac{
1-ik_{j}^{+}}{1+ik_{j}^{+}}\right) ^{n}C_{j}^{+} \nonumber \\
\fl \quad \quad { }-\frac{1}{2\pi }\int_{-\infty }^{+\infty }\left[ \mu ^{-}(n+2;-s)\left(
\frac{1-is}{1+is}\right) ^{2}-\mu ^{-}(n;-s)\right] \left( \frac{1-is}{1+is}
\right) ^{n}\rho ^{+}(s)ds.
\end{eqnarray}

\subsection{Time evolution of spectral data}

Now we want to find the time evolution of the spectral data. First, by using
(\ref{lambdak}) and (\ref{psichi}) we rewrite the auxiliary spectral problem
in terms of $\chi $ as follows
\begin{equation}
\fl \left( 1+ik\right) \chi _{1t}=\left( A_{0}-B_{0}-k^{2}B_{0}\right) \chi
_{0}+\left( 1+ik\right) \left( C_{1}-D_{1}-k^{2}D_{1}\right) \chi _{1}.
\label{chi-t}
\end{equation}
Since the $\mu ^{\pm }$'s have at large $n$ functionally different
asymptotic behaviours there exists a $\chi (t,n;k)$ satisfying (\ref{mod-eq})
and (\ref{chi-t}), and an $\Omega (t;k)$ such that
\begin{equation}
\chi (t,n;k)=\Omega (t;k)\mu ^{+}(t,n;k),
\end{equation}
that is satisfying
\begin{eqnarray}
\fl \left( 1+ik\right) \left( \Omega _{t}\Omega ^{-1}\mu _{1}^{+}+\mu
_{1t}^{+}\right) =\left( A_{0}-B_{0}-k^{2}B_{0}\right) \mu _{0}^{+} \nonumber \\
\qquad{ }+\left( 1+ik\right) \left( C_{1}-D_{1}-k^{2}D_{1}\right) \mu _{1}^{+}.
\label{Omegat}
\end{eqnarray}
We evaluate (\ref{Omegat}) separately for $n=2m$ $\rightarrow +\infty $ and
for $n=2m+1\rightarrow +\infty $ using (\ref{nu-}) and (\ref{A0})--(\ref{D1})
and recalling that $g(n)\rightarrow 2$ as $n\rightarrow \infty $. Then, we
sum up the two limits taking into account (\ref{cecd0}) getting
\begin{equation}
\Omega _{t}\Omega ^{-1}=\frac{p}{2(1+ik)}-q(1-ik)  \label{OmegaOmega-1}
\end{equation}
and, consequently, the following time-dependence for $\Omega $
\begin{equation}
\Omega (t;k)=\Omega _{0}(k)\ \exp \left[ \frac{p}{2(1+ik)}-q(1-ik)\right] t.
\end{equation}
Finally, inserting (\ref{OmegaOmega-1}) into (\ref{Omegat}) and taking the
limit $n\rightarrow -\infty $ we derive the evolution of spectral data.
Indeed, evaluating (\ref{Omegat}) on the real axis and taking into account
(\ref{nu+}) and (\ref{A0})--(\ref{D1}) we have
\begin{eqnarray}
\fl \frac{p}{2}\left( \frac{1-ik}{1+ik}\right) ^{n+1}b^{+}(k)-
q\frac{(1-ik)^{n+2}}{(1+ik)^{n}}b^{+}(k)+(1+ik)a_{t}^{+}(k) \nonumber \\
{ }+\frac{(1-ik)^{n+1}}{(1+ik)^{n}}b_{t}^{+}(k)=\frac{p}{2}\left( \frac{1-ik}{
1+ik}\right) ^{n}b^{+}(k)-q\frac{(1-ik)^{n+1}}{(1+ik)^{n-1}}b^{+}(k).\nonumber
\end{eqnarray}
Therefore
\begin{eqnarray}
a^{+}(t;k)  = a^{+}(0;k) \\
b^{+}(t;k)  = b^{+}(0;k)\exp \left[ ik\frac{p-2q(1+k^{2})}{1+k^{2}}t\right] .
\end{eqnarray}
Analogously one gets for the evolution of the normalization coefficients $
C_{j}^{+}$
\begin{equation}
C_{j}^{+}(t)=C_{j}^{+}(0)\exp \left[ ik_{j}^{+}\frac{p-2q(1+k_{j}^{+2})}{
1+k_{j}^{+2}}t\right]
\end{equation}
If $p=2q$ the evolution for $b^{+}(t;k)$ and $C_{j}^{+}(t)$ reduce to
\begin{eqnarray}
b^{+}(t;k)  = b^{+}(0;k)\exp\left[-2iq\frac{k^{3}}{1+k^{2}}t\right]  \label{b(t)} \\
C_{j}^{+}(t)  = C_{j}^{+}(0)\exp\left[-2iq\frac{k_{j}^{+3}}{1+k_{j}^{+2}}t\right]
\end{eqnarray}
which is consistent with the expected time evolution of KdV in the
continuous limit.

\section{Difference-difference equation}

We discretize now also the time and we consider the following Lax pair
\begin{eqnarray}
\psi _{20}  = \lambda \psi _{00}+g_{00}\psi _{10}  \label{principal2discrete}
\\
\psi _{11}  = \left( A_{00}+\lambda B_{00}\right) \psi _{00}+\left(
C_{10}+\lambda D_{10}\right) \psi _{10}.  \label{auxiliary2discrete}
\end{eqnarray}
If we denote by $E$ and $F$ the shift operators $Ef_{ij}=f_{i+1j}$ and $
Ff_{ij}=f_{ij+1}$ and impose to (\ref{principal2discrete}) and (\ref
{auxiliary2discrete}) the compatibility condition $E\psi _{11}=F\psi _{20}$
we get
\begin{eqnarray}
\fl \left( A_{10}+\lambda B_{10}\right) \psi _{10}+\left( C_{20}+\lambda
D_{20}\right) \left[ \lambda \psi _{00}+g_{00}\psi _{10}\right] \nonumber \\
{ }=\left( A_{-10}+\lambda B_{-10}\right) \left[ \psi _{10}-g_{-10}\psi _{00}
\right] +\lambda \left( C_{00}+\lambda D_{00}\right) \psi _{00} \nonumber \\
\qquad \qquad \qquad{ }+g_{01}\left[ \left( A_{00}+\lambda B_{00}\right) \psi _{00}+\left(
C_{10}+\lambda D_{10}\right) \psi _{10}\right]  \nonumber
\end{eqnarray}
and, then, equating the coefficients of $\psi _{00}$, $\psi _{10}$ and of
the different powers of $\lambda $ we get
\begin{eqnarray}
  D_{20}=D_{00} \\
  C_{20}-C_{00}=g_{01}B_{00}-g_{-10}B_{-10}  \label{CB} \\
  g_{01}A_{00}=g_{-10}A_{-10}  \label{gA} \\
  B_{10}-B_{-10}=g_{01}D_{10}-g_{00}D_{20}  \label{BD} \\
  -A_{10}+A_{-10} =g_{00}C_{20}-g_{01}C_{10}.  \label{AgC}
\end{eqnarray}
We choose
\begin{eqnarray}
D_{ij}  = 0 \\
B_{ij}  = b
\end{eqnarray}
and, then, this system of nonlinear recursion relations can be explicitly
solved. It is convenient to introduce the following notations
\begin{eqnarray}
\gamma _{00}  = \frac{g_{01}}{g_{00}}\frac{g_{-11}}{g_{-10}}-1 \\
G_{00}  = \left( g_{01}-g_{-10}\right) \\
\beta _{00}  = \frac{1}{g_{-20}g_{-11}}\left( \frac{g_{-20}g_{-10}}{
g_{11}g_{01}}-1\right) .
\end{eqnarray}
We get for the coefficients of the auxiliary spectral problem
\begin{eqnarray}
  \tau _{00}A_{00}+b\sigma _{00}=0 \\
  \tau _{00}\gamma _{00}C_{00}+b\alpha _{00}\sigma _{00}=b\tau _{00}G_{00}
\end{eqnarray}
where
\begin{eqnarray}
\tau _{00}  = (g_{00})^{2}\beta _{20}\gamma _{10}-(g_{01})^{2}\beta
_{10}\gamma _{20} \\
\sigma _{00}  = g_{00}\gamma _{10}G_{20}-g_{01}\gamma _{20}G_{10} \\
\alpha _{00}  = \frac{1}{g_{11}}-\frac{g_{01}}{g_{-10}g_{-20}}(1+\gamma
_{00})
\end{eqnarray}
and the nonlinear double discrete equation for $g$
\begin{eqnarray}
\fl g_{00}g_{01}\left[ \left( g_{-10}\right) ^{2}\beta _{10}\gamma _{00}-\left(
g_{-11}\right) ^{2}\beta _{00}\gamma _{10}\right] G_{20}\nonumber \\
\fl \qquad { }+(g_{01})^{2}\left[ \left( g_{-11}\right) ^{2}\beta _{00}\gamma _{20}-\left(
g_{-10}/g_{01}\right) ^{2}\left( g_{00}\right) ^{2}\beta _{20}\gamma _{00}
\right] G_{10} \nonumber \\
\fl \qquad \qquad { }+g_{-10}g_{-11}\left[ \left( g_{00}\right) ^{2}\beta _{20}\gamma
_{10}-\left( g_{01}\right) ^{2}\beta _{10}\gamma _{20}\right] G_{00}=0.
\label{double-discrete}
\end{eqnarray}
If we let
\begin{eqnarray}
  g(n,m)=g_{n}(mh) \\
  t=mh,
\end{eqnarray}
taking into account that in the limit $h\rightarrow 0$
\begin{eqnarray}
\frac{\gamma (n,m)}{h}  \rightarrow  \frac{(g_{n}g_{n-1})_{t}}{g_{n}g_{n-1}}
\\
\beta (n,m)  \rightarrow  \frac{1}{g_{n+1}g_{n}}-\frac{1}{g_{n-1}g_{n-2}} \\
G(n,m)  \rightarrow  g_{n}-g_{n-1},
\end{eqnarray}
it is easy to check that this equation in the limit of continuous time is
satisfied by the differential-difference equation (\ref{equationg}) which
contains two arbitrary constants $p$ and $q$. We expect that also in the
general case the equation (\ref{double-discrete}) could be integrated to a
lower order equation containing two arbitrary constants.

\ack The authors acknowledge A. Shabat for calling
their attention to the special merits of the discrete version of the
Schr\"{o}dinger equation used in this paper. This work was partially
supported by COFIN 2000 ``Sintesi''.

\end{document}